\documentclass[12pt]{article}

\usepackage[utf8]{inputenc}
\usepackage{amsmath}
\usepackage{amssymb}
\usepackage{amsthm}
\usepackage{blindtext}
\usepackage{mathtools}
\usepackage{cite}
\usepackage{geometry}
\usepackage{colortbl}
\usepackage{eurosym}

\usepackage{tikz}
\usepackage{caption}
\usepackage{subcaption}
\usepackage{multicol}
\usepackage{etoolbox}
\usepackage{setspace}
\usepackage{standalone}
\usepackage{multirow}
\usepackage{longtable}
\usepackage{authblk}
\usepackage{verbatim}
\usepackage{hyperref}
\hypersetup{
    colorlinks=true,
    linkcolor=blue,
    filecolor=magenta,      
    urlcolor=red,
}

\usepackage[toc,page]{appendix}

\usetikzlibrary{fit, shapes}

\tikzset{font={\fontsize{10pt}{12}\selectfont}}

\usepackage{siunitx}

\AtBeginEnvironment{quote}{\singlespace\vspace{10pt}\small}
\AtEndEnvironment{quote}{\vspace{10pt}\endsinglespace}

\begin{document}

\newcommand{\detailtexcount}[1]{%
  \immediate\write18{texcount -merge -sum -q #1.tex output.bbl > #1.wcdetail }%
  \verbatiminput{#1.wcdetail}%
}
 
\newcommand{\mb}[1]{\textcolor{cyan}{[Manuel: #1]}}
\newcommand{\sjp}[1]{\textcolor{purple}{[Stefan: #1]}}
\newcommand{\bp}[1]{\textcolor{orange}{[Bryn: #1]}}

% \detailtexcount{main}

\title{Harder, better, faster, stronger: understanding and improving the tractability of large energy system models}

\author[1]{Manuel Bröchin*}
\author[1]{Bryn Pickering}
\author[1]{Tim Tröndle}
\author[2]{Stefan Pfenninger}
\affil[1]{Institute for Environmental Decisions, Department for Environmental Systems Science, ETH Zürich, Switzerland}
\affil[2]{Faculty of Technology, Policy and Management (TPM), Delft University of Technology, Delft, The Netherlands}
\maketitle

*Corresponding author: manuel.broechin@usys.ethz.ch

\section{Abstract}
%LP is not hard, it is a solved issue. You can do everything wrong and still get the right result. But people are increasingly wanting to build larger and larger LP models to deal with the energy transition. So by applying some best practice you can dramatically increase performance and get it better, faster and stronger. Here we show you how to do it right and become a happier person and more successful energy modeller.
Energy system models based on linear programming have been growing in size with the increasing need to model renewables with high spatial and temporal detail. 
Larger models lead to high computational requirements. 
Furthermore, seemingly small changes in a model can lead to drastic differences in runtime. 
Here, we investigate measures to address this issue. 
We review the mathematical structure of a typical energy system model, and discuss issues of sparsity, degeneracy and large numerical range. 
We introduce and test a method to automatically scale models to improve numerical range. 
We test this method as well as tweaks to model formulation and solver preferences, finding that adjustments can have a substantial impact on runtime. 
In particular, the barrier method without crossover can be very fast, but affects the structure of the resulting optimal solution. 
We conclude with a range of recommendations for energy system modellers. 

Keywords: Energy system models, Scaling, Linear programming, Numerical issues, Interior-point, Simplex, Benchmark

\section{Statements and Declarations}

\subsection{Author contributions}

M.B., B.P, T.T and S.P. designed the research and developed the models used in the experiments, M.B. performed the research, analysed the data, and plotted the figures, M.B. and S.P. drafted the manuscript, M.B., B.P., T.T. and S.P. discussed and revised the manuscript.

\subsection{Funding}

The authors acknowledge funding from the European Union’s Horizon 2020 research and innovation programme under grant agreement No 837089, and the SEEDS project supported by the CHIST-ERA grant CHIST-ERA-19-CES-004, the Swiss National Science Foundation grant number 195537, the Fundação para a Ciência e Tecnologia (FCT) grant number CHIST-ERA/0005/2019, the Spanish Agencia Estatal de Investigación with grant PCI2020-120710-2, and the Estonian Research Council grant number 4-8/20/26.

\subsection{Competing Interests}

The authors have no competing interests to declare that are relevant to the content of this article.

\subsection{Software and data availability}

All experiments in this work were performed using the Calliope open-source modelling framework, accessible on Github \footnote{https://github.com/calliope-project/calliope}.
The automated scaling method was implemented on a Calliope fork, accessible on Github \footnote{https://github.com/brmanuel/calliope}.
Further, all Calliope models used in the experiments are available on Github \footnote{https://github.com/brmanuel/calliope-models}.

\section{Introduction}
Mathematical optimisation, in particular linear programming (LP), has become the method of choice for the majority of established and emerging energy system modelling tools \cite{Connolly_review_2010,Ringkjob_review_2018,Prina_Classification_2020,Chang_Trends_2021}. Renewable energy technologies need to be represented with a high spatio-temporal resolution and scope, to capture and account for the effect of their intermittency on system stability, to ensure that weather variability can be captured across years \cite{STAFFELL201865}, and to exploit the balancing effect of geographically distant weather systems \cite{Grams_Balancing_2017}. This has led to the development of ever larger models \cite{Horsch_PyPSAEur_2018,euromodel,Zappa_100_2019}. Model size increases further when using Monte-Carlo or scenario-based methods to deal with structural and parametric uncertainty \cite{Gabrielli_Robust_2019,Pickering_District_2019,DeCarolis_Modelling_2016,Lombardi_Policy_2020}.

\cite{Pfenninger_Energy_2014} identified tractability as one key challenge for energy system optimisation models. Most commonly, temporal resolution is reduced \cite{euromodel,Majewski_Robust_2017,Burandt_Decarbonizing_2019} or time periods with similar features are clustered and represented only by a subset of `typical days' \cite{Pickering_District_2019,Mavromatidis_Comparison_2018,Babonneau_ETEMSG_2017}. Advancements in the field of time-series aggregation include pre-selection of `critical' days prior to clustering \cite{Pfenninger_Dealing_2017}, identification of k-mediods as the most reliable aggregation algorithm \cite{Schutz_Comparison_2018,Kotzur_Impact_2018}, and separation of storage into multiple decision variables, to allow information to be linked within and between clustered periods \cite{Kotzur_Time_2018,Gabrielli_Optimal_2018}. \cite{Hoffmann_Review_2020} summarised these methods, finding that existing feature aggregation using k-means, k-mediods, or hierarchical clustering is still the `state-of-the-art'. However, they reiterated warnings made in several previous studies that time-series aggregation alters model results, both qualitatively and quantitatively, and should therefore be used with caution \cite{Pfenninger_Dealing_2017,Kotzur_Impact_2018}. Recent work \cite{beam_me} tries to improve tractability of energy system models on the algorithmic layer: they develop a parallel algorithm that exploits the inherent structure of energy system models to break up the problem into weakly related subproblems that can then be solved individually.

Similarly to \cite{beam_me}, we take a step back from application-specific methods to reduce model complexity. Instead, we examine how the structure of the underlying optimisation problem affects its tractability, and empirically examine a range of options to improve tractability for typical energy system models. We also develop and test a method to facilitate tractability, in which we automatically scale the parameters of an energy system model. To do so, we draw on literature from the broader field of operations research \cite{guidelines_klotz, ref_scaling_12, ref_scaling_75}. The paper proceeds as follows. First, we discuss the mathematical properties of a typical high-resolution energy system model and compare the characteristics of the two main solution methods: simplex and barrier (interior point). We identify scaling of the optimisation problem as a particular area of concern for the performance of these methods and introduce a method to automatically scale an energy system model. We then proceed with a series of systematic experiments investigating different solver configurations, examining whether we can save computational effort by settling for a potentially less accurate solution, and testing the performance of our scaling method. We then discuss what general guidelines for energy system modelling we can draw from these experiments. We use the open-source Calliope modelling tool \cite{Pfenninger2018} for our experiments.

\section{Background} \label{sec_background}

A high-level view of the problem we want to solve is the following: Given the geographically distributed consumption and production of "goods" such as electricity or heat (from now on referred to as `carriers'), a network over which these carriers can be transferred, and the ability to store and convert between carriers, we want to find the cheapest way to allocate these carriers over time and between locations such that all demand is met. In other words, we want to solve a \textit{minimum-cost flow problem} with multiple commodities. In practice we find that network flow problems are not expressive enough to formulate realistic energy system models. In particular, imposing policy-based constraints on carrier consumption and production such as minimum shares of renewable generation, mean that energy system models cannot be described just as network flow problems. Instead, we must consider them as generic linear programs (LPs).~\footnote{Many energy system models additionally feature integrality constraints, i.e. they require certain variables to be integers in the solution. Ordinary LPs cannot express such constraints and a mixed-integer LP (MILP) formulation must be chosen instead. We ignore MILP specific issues in this work. However, since MILP solvers need to solve a "relaxed" version of the problem without integrality constraints as a subproblem, everything in this work should also apply to MILP problems.}

\subsection{LP problems and solution methods} \label{sec_properties}
LPs are a very general framework of mathematical optimization \cite{vanderbei2015linear}.
The general form of an LP is
\begin{align} \label{eq:lp}
\begin{split}
    \text{minimize}\ &c^\intercal x \\
    \text{s.t.}\ &Ax = b \\
    &0 \leq x
\end{split}
\end{align}
The linear objective function $x \longmapsto c^\intercal x$ is to be minimized by choosing suitable variable values $x \in \mathbb{R}_+^n$.
The constraint matrix $A \in \mathbb{R}^{m\times n}$ and right-hand side vector $b \in \mathbb{R}^m$ together constrain the choice of variables $x$ with $m$ linear constraints.
A positive vector $x \in \mathbb{R}_+^n$ is called \emph{feasible} if it satisfies all $m$ constraints. 
A set of $m$ independent columns of $A$ induces a basis $A_B$ of the column space of $A$. A feasible point $x$ can be obtained by solving $A x = b$ while setting all variables not corresponding to basis columns to $0$. We call such a solution $x$ a basic feasible solution of equation \ref{eq:lp}. %Note that we can assume without loss of generality that the rank of $A$ is $m$, else $Ax = b$ either has no solution at all or $A$ contains linearly dependent rows. Both cases can easily be detected and dealt with. 
Geometrically, a basic feasible solution is an extreme point (a vertex) of the feasible region. 

Formulating an energy system in analogy to cost-minimal flow problems as an LP is straightforward:
$x_i$ is the variable holding the amount of flow through the $i$-th edge. Row $j$ of $A$ contains a non-zero entry at index $i$ which is $1$ or $-1$ if the $i$-th edge is an incoming or outgoing edge of node $j$ respectively. $b_j$ is the flow-demand of node $j$. $c_i$ is the cost incurred by one unit of flow at edge $i$. Additionally, edge capacities and many constraints that we find in energy system models, such as efficiencies of carrier conversion or complex policies governing the operation of certain plants, can either be expressed directly or approximated by linear constraints.

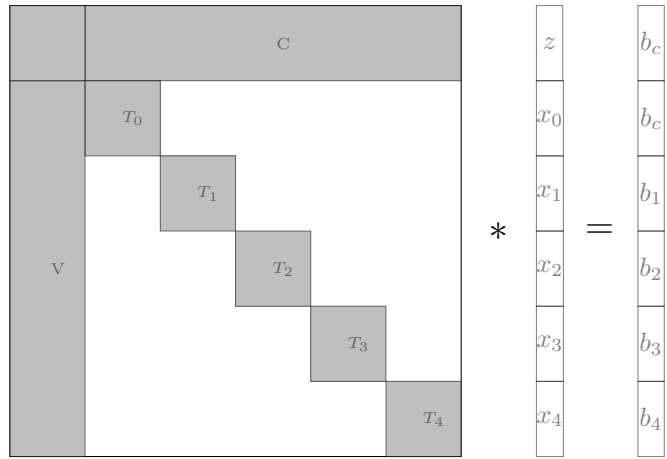
\begin{figure}[t] 
\centering
\begin{tikzpicture}
\tikzstyle{node} = [draw=black, rectangle, opacity=.5, text opacity=1, fill=gray, inner sep=0, font=\tiny]
\tikzstyle{var} = [draw=black, rectangle, opacity=.5, inner sep=0, minimum height=1cm, minimum width = 0.35cm, anchor=south west]

% matrix
\node [draw=black, rectangle, minimum width=6cm, minimum height = 6cm, anchor=south west] at (0,1) {};

\node  [node, minimum width=5cm, minimum height = 1cm, anchor=south west] at (1,6) {C};
\node  [node, minimum height=5cm, minimum width = 1cm, anchor=south west] at (0,1) {V};
\node  [node, minimum height=1cm, minimum width = 1cm, anchor=south west] at (0,6) {};

\node  [node, minimum height=1cm, minimum width = 1cm, anchor=south west] at (1,5) {$T_0$};
\node  [node, minimum height=1cm, minimum width = 1cm, anchor=south west] at (2,4) {$T_1$};
\node  [node, minimum height=1cm, minimum width = 1cm, anchor=south west] at (3,3) {$T_2$};
\node  [node, minimum height=1cm, minimum width = 1cm, anchor=south west] at (4,2) {$T_3$};
\node  [node, minimum height=1cm, minimum width = 1cm, anchor=south west] at (5,1) {$T_4$};

% mult
\node [draw=none] at (6.5,4) {\large{$*$}};
\node [draw=none] at (7.85,4) {\large{$=$}};

% vector
%\node [draw=black, rectangle, minimum width=0.35cm, minimum height = 6cm, anchor=south west] at (7,1) {};
\foreach \x\v\b in {6/$z$/$b_c$, 5/$x_0$/$b_c$, 4/$x_1$/$b_1$, 3/$x_2$/$b_2$, 2/$x_3$/$b_3$, 1/$x_4$/$b_4$}
{
    \node  [var] at (7,\x) {\v};
    \node  [var] at (8.35,\x) {\b};
}   

\end{tikzpicture}
\caption{$A\cdot x = b$ in block-matrix notation showing the typical sparsity pattern of the constraint matrix $A$ of an energy system model. Variables $x_i$ are specific to timestep $i$ and are constrained using parameters $\tau_i$ and $b_i$. $cx = b_c$ are constraints containing dependencies across time steps. Parameters $v$ and variables $z$ model time-independent quantities.}
\label{fig:arrowhead} 
\end{figure}

As discussed above, energy system models consider the allocation of resources in time and space. Modelling time requires its discretisation into timesteps, which leads to a very particular structure of the constraint matrix $A$, depicted in Figure \ref{fig:arrowhead}.
This structure is often referred to as \emph{arrowhead} structure. Specialized LP solution methods which try to exploit this structure to speed up the solution process are in development \cite{ref_parallel_ip}.

At the moment, the two families of algorithms most widely used to solve energy system models are the simplex algorithms and barrier or interior-point methods, of which many variants exist. These two approaches are available both in open-source software packages such as Coin-OR \cite{Coinor2003} or commercial ones such as Gurobi \cite{Gurobi2021}. In the following we sketch some important characteristics of both simplex and barrier methods based on the treatment in \cite{guidelines_klotz}. Table \ref{tab:algo_characteristics} lists the main points for both methods side-by-side.

Although the two algorithms solve the same problem, they operate in entirely different ways. Simplex maintains and updates a basic feasible solution $x_i$ in every step $i$ of the algorithm. To find $x_{i+1}$, simplex chooses one of the neighboring vertices of $x_i$ that improves the objective value. This jump from $x_i$ to $x_{i+1}$ is performed algebraically by exchanging a column in the current basis of $A$. Convexity of the feasible region and linearity of the objective function imply that an $x_i$ which cannot be improved in this way is optimal. 

Barrier maintains an interior point $x$ of the feasible region. In every step of the algorithm, barrier finds a vector along which $x$ can be improved with respect to $c$, and takes a step along this direction without leaving the feasible region. With $x$ becoming closer to optimal, $x$ will also move towards the boundary of the feasible region. Once the update step of $x$ is shorter than some threshold, barrier will consider $x$ to be close enough to optimal value and return it as the solution. 

Simplex always returns a basic feasible solution while barrier in general returns an interior point of the feasible region. This has several implications: 

First, an interior point is, strictly speaking, not an optimal solution to an LP. However, the "gap" can in theory be made arbitrarily small. Moreover, optimality of a basic feasible solution can also only be determined with a certain limited precision. Thus in practice, solvers typically allow controlling the precision of both methods. Neither "optimal" solutions returned by simplex nor by barrier are generally better, but the precision of the solution depends on the solver configuration.

Second, there are cases where a basic feasible solution is needed: (1) when solving a MILP problem, and (2) if the obtained solution is to be used to warm-start subsequent solver runs on the same (or a slightly different) model.

Third, an interior solution has potentially many more variables away from their bounds than a basic solution. Since variables representing real-life quantities often have a lower bound of $0$, this means that solutions returned by barrier usually have many more non-zeros than basic feasible solutions; potentially these non-zeros are unrealistically small for the considered problem. A solution with many non-zeros may be harder to interpret and may thus be undesirable. 

To rectify these downsides of interior solutions, a dedicated crossover method can be used to find an optimal basic feasible solution from an optimal interior solution \cite{ref_recovering_basis_bixby}.

\begin{table}[t!]
    \footnotesize
    \centering
    \begin{tabular}{|p{2cm}|p{6cm}|p{6cm}|}
        \hline 
         &\textbf{Primal simplex} & \textbf{Barrier (Primal Affine Scaling)} \\
         \hline 
         State 
         & Maintains a basis of $A$ and a corresponding basic feasible solution
         & Maintains a non-basic (interior) point $x$ satisfying $A x = b$ \\
         \hline
         Update 
         & Performs row operations on $A$ to update the basis in each step (pivot operation) 
         & Computes a vector along which the objective is improved, projects it onto the nullspace of $A$ and updates $x$ along this vector while maintaining $x > 0$ \\
         \hline 
         Main computation
         & Main computational cost is solving $m\times m$ linear systems of equations for the pivot step
         & Main computational cost is projecting the update vector onto the nullspace of $A$. This entails inverting the $m\times m$ matrix $\hat{A}\hat{A}^\intercal$ in every step where $\hat{A}$ is a scaled version of $A$. \\
         \hline
         Termination condition 
         & Terminates if no basis update can improve the objective value over the current basic feasible solution
         & Terminates if the last update step was shorter than some threshold \\
         \hline 
         Return value 
         & Returns a basic feasible solution
         & Returns an interior point that is close to the boundary of the feasible region (we will denote this an interior solution). A crossover method can be added to retrieve an optimal basic feasible solution from an interior solution \\ 
         \hline 
    \end{tabular}
    \caption{Comparison of major characteristics of simplex and barrier methods}
    \label{tab:algo_characteristics}
\end{table}

Neither of the two methods is generally faster than the other. However, the factorization step of $\hat{A}\hat{A}^\intercal$ in barrier is more amenable to parallelization than the pivot operations of simplex \cite{guidelines_klotz}. Nevertheless, the most suitable algorithm depends on the structure of the LP at hand \cite{Tomlin1989, guidelines_klotz}. 
Two LPs of the same size can lead to vastly different solution times for both methods, simplex and barrier. This is mostly due to two reasons: numerical solvers can recognize and exploit problems with special structures and they are susceptible to numerical problems. Three aspects are particularly influential: sparsity (structure), degeneracy (structure), and the numerical range.

The first aspect impacting the solution time is sparsity. Energy system models typically lead to LPs that are highly sparse. This follows directly from the arrowhead structure depicted in Figure \ref{fig:arrowhead}: the number of zero entries scales quadratically in the number of time steps, whereas the number of nonzero entries is linear in the number of time steps. A typical model might consider a full year at a granularity of 1 hour time steps, causing most entries of the matrix to be zero. In general, sparsity is a desirable property of a linear program because both simplex and barrier solvers have ways of profiting from sparsity in the constraint matrix $A$.
Interestingly, barrier is very sensitive to the sparsity-pattern of $A$ while it matters almost not at all to simplex. The reason for this is that for barrier the product $AA^\intercal$ needs to be sparse. Note that if $A$ has some dense columns, then $AA^\intercal$ will be dense too. On the other hand, if $A$ has no dense columns but some dense rows, then $AA^\intercal$ will be sparse. Simplex, on the other hand, relies on the LU decompositions of bases of $A$ being sparse. \cite{luce2009factorization} observe that a large part of most bases of $A$ can be permuted to triangular form. Thus, for most bases only a very small part actually needs to be factorized, the largest part corresponds directly to the basis itself and thus the resulting LU decomposition will be almost as sparse as the original basis. 

The second aspect that influences solution time is degeneracy. 
Degeneracy in the context of linear programming means that multiple bases of the constraint matrix $A$ lead to the same basic feasible solution. 
Standard minimum-cost flow problems are often inherently degenerate \cite{network_flows, Tomlin1989}.
Degeneracy can pose significant problems to the simplex method: because multiple bases lead to the same solution, simplex updates may fail to improve the objective and stall progress. Computational studies have shown that up to 90\% of all pivot operations of the simplex method on min-cost flow problems can be degenerate \cite{network_flows}. 
Barrier methods maintain an interior point of the feasible region and never actually encounter a basic feasible solution. Degeneracy thus matters less to the operation of barrier methods \cite{guidelines_klotz}. 

The third and final aspect is a large numerical range of problems, given by the absolute values of coefficients in the constraint matrix, the right-hand side and the objective function. Energy system models often have a very large numerical range. This is a consequence of two facts: Firstly, the input data of energy system models correspond to different physical quantities (e.g. energy, area, cost). Choosing inappropriate combinations of units to represent these quantities will lead to large numerical ranges. Secondly, even within quantities of the same unit we might encounter large numerical ranges, for instance, between operating costs and investment costs for new generation infrastructure, or even just between the operating costs of very different technologies like photovoltaics and gas-fired power generation. A large numerical range can result in increased solution times and even lead to non-convergence in extreme cases. It can furthermore lead to loss of precision due to round-off errors. To address these issues, it is possible to scale the linear problem before solving it. 

\subsection{Scaling} \label{sec_scaling}

Choosing positive scaling factors $r_1, \dots, r_m, s_1, \dots, s_n \in \mathbb{R}_+$ we can scale each row $i$ of $A$ by $r_i$ and each column $j$ of $A$ by $s_j$ by multiplying $A$ from left and right with the diagonal matrices $R := \text{diag}(r_1, \dots, r_m)$ and $S := \text{diag}(s_1, \dots, s_n)$.
\begin{equation} \label{eq:lp:unscaled}
    \min \{c^\intercal x \ |\ x \geq 0, Ax = b\}
\end{equation}
\begin{equation} \label{eq:lp:scaled}
    \min \{c^\intercal S x\ |\ x \geq 0, RASx = Rb\}
\end{equation}
Perhaps unintuitively, the original problem \ref{eq:lp:unscaled} is equivalent to problem  \ref{eq:lp:scaled} which is scaled with $R$ and $S$. A short proof for this is in the Appendix \ref{sec_appendix_proof}.
While being equivalent, problem \ref{eq:lp:unscaled} may have better numerical properties than the original LP. In the following we consider the numerical range and the condition number of $A$ as two factors that decide whether an LP has good or bad numerical properties.

The numerical range $\frac{\max_{x \in A, b, c} |x|}{\min_{x \in A, b, c} |x|}$ of an LP impacts the performance of the solution methods. The reason for this is that most solvers use absolute tolerances to compare numbers. For instance, a central parameter of the Gurobi solver is the feasibility tolerance $\tau_f$, set by default to $\tau_f = 10^{-6}$. To check if a point $x$ satisfies some constraint $A_i x \leq b_i$, Gurobi checks whether $A_ix - b_i \leq \tau_f$. If such computations are at the order of $10^{10}$, then the relative error of $10^{-16}$ induced by floating point arithmetic is at the order of the tolerance $\tau_f$. In other words, feasibility can no longer be reliably computed. Similarly, if some computation is at the order of $10^{-6}$, the result will likely be meaningless to Gurobi. If our model now includes both very large and very small numbers, it is likely that both types of problems appear during the solution process. It is thus desirable to limit the numerical range in our model formulation. Note that this applies equally to both simplex and barrier methods.

A different notion is the condition number $\kappa(M) := \lVert M \rVert \lVert M^{-1} \rVert$ of a matrix $M$. Intuitively, the condition number of a matrix measures the effect of small rounding errors when solving the linear system of equations associated with the matrix: For a matrix with small condition number, the rounding errors will barely impact the solution. For a matrix with large condition number, the effect of the rounding error can be significant. Note that both simplex and barrier rely heavily on solving linear systems of equations. In the case of simplex, these matrices are the square submatrices of $A$ corresponding to the visited bases of $A$. In the case of barrier, the important matrices are $\hat{A}\hat{A}^\intercal$ that are inverted at each step (see Table \ref{tab:algo_characteristics}).

Both factors, the numerical range and the condition number thus impact the accuracy of the computations performed during both simplex and barrier and both are affected by scaling.
The relationship between numerical range and condition number, however, is complicated: improving one may make the other one worse. We elaborate more on this in Appendix \ref{sec_appendix_condition_number}.
The accuracy of computations impacts the performance of these algorithms in multiple ways. First, these algorithms make decisions based on numerical computations. For instance, simplex chooses the next basic feasible solution based on some metric which is computed numerically. Large errors in the computation of this metric may cause simplex to make a suboptimal choice and increase the number of steps necessary to reach optimality. Second, to avoid the previous issue, solvers have ways to detect and deal with loss of precision. One way is switching to a numeric data type with higher precision. These types incur a higher computational cost for each operation and thus cause overall slowdown of the algorithm (a secondary impact would additionally be the higher memory cost of higher-precision data types, which in itself can be significant for very large models).

\subsection{An automated scaling method}

Both numerical range and condition number of a matrix are affected by scaling rows and columns of the matrix. 
While reducing numerical range with scaling is straightforward, reducing the condition number of all relevant matrices that are encountered during simplex and barrier is not.
In the case of simplex, \cite{ref_scaling_12} compares many different scaling methods and their effect on the condition number. For certain problems, scaling can lead to an increase of the condition number averaged over all basis matrices encountered during a simplex run. 

We have developed an automatic scaling method that focuses on minimizing the numerical range of the input data, which ignores the issue of the condition number (we refer to this method as "auto-scaling" from here one).
The basic idea is to group input values by type (such as area, cost, energy, etc.) and scale all values in one group with the same factor.
The assumption is that values of the same type will often be contained within an acceptable range.
This approach corresponds to choosing suitable units (such as $m^2$, dollars, kWh, etc.) for each type of quantity.
Apart from being a natural approach to reducing numerical range, grouping values together this way considerably reduces the search space for good scaling factors.
We formulate the choice of good scaling factors as a small, auxiliary optimization problem. Details are given in the appendix \ref{sec_method_scaling}.

\section{Experimental Procedures} \label{subsec_methods}
We consider models generated by the Calliope modelling framework \cite{Pfenninger2018} and want to investigate (1) which algorithm solves these problems in the least amount of time, (2) what the trade-offs when choosing between a basic feasible solution and an interior solution are, and (3) what the impact of our scaling approach on the solution time is. The models used for the experiments are specified in Table \ref{tab:models}.

\begin{table}[ht!]
    \footnotesize
    \centering
    \begin{tabular}{|l|l|l|}
    \hline
     Name & Model & Citation\\
     \hline
     Euro & 34-zone LP Euro-Calliope model\footnotemark & \cite{euromodel} \\
     UK & 30-zone LP UK-Calliope\footnotemark & \cite{ukmodel}\\
     Bangalore & 10-zone MILP Bangalore-Calliope\footnotemark & \cite{bangmodel} \\
     BangaloreLP & Same as Bangalore but without integrality constraints & \cite{bangmodel} \\
     \hline
    \end{tabular}
    \caption{Calliope models used in experiments. Note that the Euro model by default includes manual scaling to improve performance; for the experiments here we undo this manual scaling.}
    \label{tab:models}
\end{table}
\footnotetext[2]{https://github.com/calliope-project/euro-calliope}
\footnotetext[3]{https://github.com/calliope-project/uk-calliope}
\footnotetext[4]{https://github.com/brynpickering/bangalore-calliope}

All models are implemented with the Calliope modelling framework. We will often add a postfix to model names to indicate the time range, e.g. Euro\_15d and Euro\_5m refer to the Euro-Calliope model run over a 15-day and a 5-month time range respectively, starting from January 1, always at the time step resolution of 1 hour. Note that the Bangalore model is set in a leap year while Euro and UK are not. Because we set the time ranges by date, the selected time range in the Bangalore model is often one day longer than in the other two models. 

We run all benchmarks on the Euler cluster of ETH Zurich. All experiments where solution times are measured are performed on the same setup: compute nodes with two 18-core Intel Xeon Gold 6150 processors (2.7-3.7 GHz) with 192 GB of DDR4 memory clocked at 2666 MHz. Each experiment was run alone on a full node, i.e. taking up all 36 cores, to prevent competing processes from other cluster users from affecting the solution time. Note, however, that the number of cores does not correspond to the number of software threads. Except where explicitly noted we set the number of software threads for the solvers to 4.  To diagnose numerical issues in our models we closely analyze the logs produced by the solver. This methodology is showcased for instance in \cite{guidelines_klotz}. 

\paragraph{Solver choice.}
We compare the suitability of the commercial Gurobi solver and the open-source Coin-OR Clp/CBC solver for our problems. We compare both simplex and barrier methods of these two solvers. For Gurobi specifically, we also investigate whether we can improve solution times through solver parameters. Other authors have performed more thorough solver comparisons on a wider class of problems \footnote{http://plato.asu.edu/guide.html}.

The authors of \cite{performance_variability} observe that the performance of MILP solvers is non-deterministic and is often subject to large variation due to seemingly unimportant changes: different machines, compilers, and libraries can all lead to vastly different performance. We observe the same behaviour for Gurobi's LP solvers, mostly for the crossover phase (including final simplex cleanup). In order to account for this variability in our benchmarks, we perform each measurement several times only changing the random seed value of Gurobi. For each benchmark we indicate the number of repetitions, the average and min/max solution times.

\paragraph{Interior vs. basic solution.}
As discussed in Section \ref{sec_properties}, the barrier method run by itself returns an interior solution which is approximately optimal, whereas the simplex method returns a basic feasible solution. When using the barrier method, it is possible to run the crossover method to find a basic feasible solution, taking the interior solution as the initial value. We consider up- and downsides of using an interior solution as compared to obtaining a basic solution. 

We consider two questions: (1) how quickly we can obtain both interior and basic feasible solutions, and (2) how "good" either solution is. To answer (1) we perform benchmarks for each of the solution methods and on several models. To answer (2) we analyze the returned solution in terms of several relevant metrics: solution convergence, objective function value, and the fraction of non-zeros in the solution. We then discuss how interior and basic solutions can imply different policy decisions. 

When interpreting the solution to a model we are mostly interested in the following Calliope decision variables:
\begin{itemize}
    \item \texttt{capacity}: indicates for each location and technology the installed capacity (from now on denoted \textit{cap}).
    \item \texttt{carrier\_production}:  indicates for each time step, location and technology, how much of a given carrier is produced (from now on denoted \textit{prod}).
    \item \texttt{systemwide\_levelised\_cost}: indicates the total per-unit cost paid for each carrier across the entire energy system (from now on denoted \textit{lcoe}).
\end{itemize} 

\paragraph{Scaling.} 
As discussed in Sections \ref{sec_properties} and \ref{sec_scaling}, energy system models often have large numerical ranges and it is desirable to apply scaling to avoid numerical issues. The models we examine are often numerically problematic, and we investigate whether our auto-scaling method as developed in Section \ref{sec_method_scaling} can effectively mitigate these problems. To do so, we benchmark the solution time of various models with and without auto-scaling and quantify the effects of scaling.

\section{Results} \label{sec_results}

\subsection{Solver choice} \label{sec_solver_choice}

In this comparison, we only consider pure LP models without integrality constraints, using the Gurobi and Coin-OR solvers, for primal simplex, dual simplex, and barrier method with and without crossover. For each model we create instances of different sizes by varying the number of time steps. The shortest time frames are of little practical relevance but are included here to give a more complete picture of solver capabilities.

We aim at configuring each solver and algorithm equivalently despite the inherent differences in configurability between Coin-OR and Gurobi's solvers: in particular, we instruct each solver to use 4 threads and we set optimality and feasibility tolerances to $10^{-5}$ (changing from the $10^{-6}$ default in Gurobi). 
We group our observations in the following paragraphs roughly by solver.

\begin{table}
\centering
\footnotesize
\begin{tabular}{|l|r|r|r|r|r|r|r|r|}
\hline
 & \multicolumn{4}{c|}{\textbf{Coin-OR BC}} & \multicolumn{4}{c|}{\textbf{Gurobi}} \\
 \hline
\textbf{Model} & \textbf{Bar} & \textbf{Crossover} & \textbf{Dual} & \textbf{Primal} & \textbf{Bar} & \textbf{Crossover} & \textbf{Dual} & \textbf{Primal} \\
\hline
BangaloreLP\_2d & 52 &54 &10 &9 &9 &10 &8 &9 \\
BangaloreLP\_3d & 579 &570 &14 &12 &11 &13 &11 &14 \\ 
BangaloreLP\_5d & 6017 &6019 &16 &19 &15 &18 &15 &18 \\
BangaloreLP\_11d &  $-^{(2)}$ & $-^{(2)}$ & 31 &64 &30 &31 &28 &49 \\
BangaloreLP\_31d &  $-^{(1)}$ & $-^{(1)}$ &115 &211 &79 &89 &80 &219 \\
BangaloreLP\_62d & $-^{(1)}$ & $-^{(1)}$ &328 &300 &164 &234 &225 &784 \\
BangaloreLP\_182d & $-^{(1)}$ & $-^{(1)}$ &2727 &2348 &588 &1509 &1395 & $-^{(2)}$ \\
\hline 
Euro\_2d & 70 &73 &68 &97 &63 &62 &67 &90 \\
Euro\_3d & 128 &131 &80 &153 &73 &72 &79 &158 \\
Euro\_5d & 834 &837 &128 &360 &93 &96 &121 &401 \\
Euro\_11d & $-^{(2)}$ & $-^{(2)}$ & 493 &2668 &161 &175 &441 &1547 \\
Euro\_31d &  $-^{(2)}$ & $-^{(2)}$ & $-^{(2)}$ & $-^{(2)}$ &394 &485 &5819 & $-^{(2)}$ \\
Euro\_61d &   $-^{(1)}$ &  $-^{(1)}$ & $-^{(2)}$ & $-^{(2)}$ &771 &1526 & $-^{(2)}$ &  $-^{(2)}$ \\
Euro\_181d &  $-^{(1)}$ &  $-^{(1)}$ &  $-^{(2)}$ &  $-^{(2)}$ & 2564 &8026 &  $-^{(2)}$ & $-^{(2)}$ \\
\hline 
UK\_2d & 121 &121 &19 &20 &22 &21 &22 &21 \\
UK\_3d &  475 &469 &25 &25 &28 &28 &28 &29 \\
UK\_5d & 4054 &4056 &39 &39 &42 &42 &43 &43 \\
UK\_11d &   $-^{(2)}$ & $-^{(2)}$ & 86 &89 &83 &85 &91 &87 \\
UK\_31d & $-^{(2)}$ & $-^{(2)}$ & 342 &445 &229 &230 &273 &245 \\
UK\_61d &  $-^{(3)}$ & $-^{(3)}$ & 1026 &1622 &447 &451 &708 &615 \\ 
UK\_181d &  $-^{(3)}$ & $-^{(3)}$ &  $-^{(2)}$ & $-^{(2)}$ & 1400 &1414 &4257 &2773 \\
\hline
\end{tabular}
\caption{Solution times in seconds of different solvers on different model instances. \textbf{Bar} is barrier method, \textbf{Crossover} is barrier+crossover, \textbf{Primal} and \textbf{Dual} are  primal and dual simplex, respectively. Each solution time is the average of two runs. Runs that did not terminate successfully within 4 hours and with 72GB of RAM are indicated with $-$. In those cases, we indicate the returned error: $^{(1)}$ wrong algorithm, $^{(2)}$ timeout, $^{(3)}$ solver error. See the text for more detailed information on what these errors mean.}
\label{tab:solver_comp}
\end{table}

\paragraph{Coin-OR barrier}
As an important caveat we must note that the barrier implementation of Coin-OR is meant as a baseline implementation which the user is supposed to extend with problem specific implementations. In particular, it is stated on the Clp homepage\footnote{https://www.coin-or.org/Clp/faq.html} that "the sparse factorization [of barrier] requires a good ordering algorithm, which the user is expected to provide (perhaps a better factorization code as well)."
We nevertheless use this baseline implementation for our comparison, and perhaps unsurprisingly, Coin-OR's default barrier algorithm struggles with all but the smallest problems (Table \ref{tab:solver_comp}). After a certain size Coin-OR fails to perform a single barrier iteration and either times out (error condition $^{(2)}$) or chooses to run the simplex method instead of barrier (error condition $^{(1)}$). Also for small model sizes, barrier of Coin-OR is not competitive with any of the other algorithms. 

\paragraph{Primal and dual simplex}
Primal and dual simplex implementations of Gurobi and Coin-OR each seem to be comparable in their capabilities but Gurobi is generally slightly better (Table \ref{tab:solver_comp}). Both dual simplex implementations solve almost the same set of problem instances, each with comparable solution times. Similarly, both primal simplex implementations solve almost the same set of problems with comparable solution times. Generally, dual simplex seems to be at least as good as primal simplex (with some exceptions for Gurobi, for instance UK\_181d).

\paragraph{Gurobi Barrier methods}
Gurobi's barrier method with and without crossover solves all problem instances and more instances than all other solvers (Table \ref{tab:solver_comp}). Except for the smallest problem instances, barrier methods have the shortest solution times. Turning on crossover has an unpredictable effect on runtims: in some cases solution time is barely affected (e.g. UK\_181d), in other cases the solution time triples (e.g. BangaloreLP\_182d and Euro\_181d).

\subsection{To crossover or not}\label{sec_interior}
We observe that the barrier method often solves models much faster than simplex methods do, but that that the crossover step can have a significant impact on the total solution time. It is tempting to say that disabling crossover is therefore an easy way to dramatically improve solution times. However, time to solution alone is not sufficient to choose a method because the solutions returned by barrier and those returned by simplex or barrier+crossover are inherently different. In this section we thus compare the interior solutions obtained by running barrier without crossover with the basic feasible solution obtained by running either the simplex method or barrier+crossover. 

Table \ref{tab:inter_vs_basic} compares solutions to various Calliope models that were obtained using barrier and barrier+crossover. The models considered in Table \ref{tab:inter_vs_basic} are all slight variations of the models listed in Table \ref{tab:models}. These variations were obtained by varying costs of certain technologies and, in the case of the Euro model, by reducing the size of the network. The concrete modifications performed are described in Appendix \ref{appendix:model:variations}. Solutions are compared with respect to the objective value obtained, the time it took to find the solution and the fraction of non-zero values in the solution. In this section we consider all numbers with absolute value $\leq 10^{-10}$ to be zero. For the objective value we additionally list the signed relative error $\varepsilon$ of the interior solution as compared with the basic feasible solution
\begin{equation}
    \varepsilon := \frac{\text{inter} - \text{basic}}{\text{basic}}
\end{equation}
First, we note that $\text{Bangalore}^2\text{\_181d}$ was not successfully solved by the barrier method alone. 
Closer investigation shows that this happens consistently for this model instance while it never happens for the closely related model $\text{Bangalore}^1\text{\_181d}$.
The two models are identical apart from slightly different costs associated with technologies. As described in Appendix \ref{appendix:model:variations}, $\text{Bangalore}^2\text{\_181d}$ scales the cost contribution of carbon by $0.365$ compared to the original model and by a factor of 26 compared to $\text{Bangalore}^1\text{\_181d}$. Moreover,  the Bangalore model associates dummy carbon costs with all technologies, whose sole purpose it is to prevent the solver from allocating unused capacities. The use of these technologies is otherwise unbounded. 
The combined effect of these modelling decisions is that $\text{Bangalore}^2\text{\_181d}$ contains a set of unbounded variables whose cost contribution is almost zero (and lower than in other instances of this model).
This formulation geometrically leads to unbounded faces that are almost ``flat'', i.e. even distant points on the face have almost the same cost. 
It is known that unbounded optimal faces may lead to numerical issues for barrier methods~\cite{cplex}.
Indeed, associating slightly higher costs to electricity\_transmission in $\text{Bangalore}^2\text{\_181d}$ resolves the issue entirely: barrier consistently solves the modified problem to optimality. 
This case illustrates that barrier can be more susceptible to numerical issues than barrier+crossover. 
We will return to this insight in Section \ref{sec:discussion} where we develop guidelines for model formulation.

Next, we observe that the interior solution is generally slightly worse than the basic feasible solution. However, the differences are negligible in all cases. Moreover, we recall that a better approximation can be obtained by tightening the barrier convergence tolerance parameter. 

Barrier+crossover always performs additional work compared to just barrier, thus obtaining a basic feasible solution takes longer than obtaining an interior solution. A counter-example to this intuitive rule is the $\text{UK}^1\text{\_180d}$ where obtaining an interior solution takes longer than obtaining a basic feasible solution. This is most likely due to solution time variability as discussed in Section \ref{subsec_methods}. 

\begin{table}[ht!]
\centering
\footnotesize
\begin{tabular}{|l|c|c|c|r|r|l|l|}
\hline
 & \multicolumn{3}{c|}{\textbf{Objective}} & \multicolumn{2}{c|}{\textbf{Time}} & \multicolumn{2}{c|}{\textbf{Non-zeros}} \\
 \hline
\textbf{Model} & \textbf{basic} & \textbf{inter} & \textbf{$\varepsilon$} & \textbf{basic} & \textbf{inter} & \textbf{basic} & \textbf{inter}  \\
\hline 
$\text{Euro}^1\text{\_180d}$ & 7.34e+10 & 7.34e+10 & 5e-07 & 4008 & 1869 & 0.27 & 0.62 \\
$\text{Euro}^2\text{\_180d}$ & 6.06e+10 & 6.06e+10 & 1e-08 & 3860 & 2313 & 0.29 & 0.58 \\
$\text{Euro}^3\text{\_180d}$ & 2.20e+10 & 2.20e+10 & 2e-10 & 65 & 63 & 0.42 & 0.6 \\
$\text{Euro}^4\text{\_180d}$ & 2.19e+10 & 2.19e+10 & 5e-16 & 56 & 54 & 0.43 & 0.56 \\
$\text{Euro}^5\text{\_180d}$ & 2.17e+10 & 2.17e+10 & 9e-11 & 57 & 52 & 0.43 & 0.6 \\
$\text{Euro}^6\text{\_180d}$ & 2.20e+10 & 2.20e+10 & 4e-12 & 52 & 50 & 0.42 & 0.58 \\
$\text{UK}^1\text{\_180d}$ & 1.24e+10 & 1.24e+10 & 3e-11 & 2038 & 2274 & 0.25 & 0.48 \\
$\text{UK}^2\text{\_180d}$ & 1.70e+10 & 1.70e+10 & 6e-08 & 2788 & 2215 & 0.26 & 0.52 \\
$\text{UK}^3\text{\_180d}$ & 1.27e+10 & 1.27e+10 & 2e-09 & 5081 & 4105 & 0.24 & 0.5 \\
$\text{UK}^4\text{\_180d}$ & 1.29e+10 & 1.29e+10 & 3e-09 & 3399 & 2239 & 0.25 & 0.5 \\
$\text{UK}^5\text{\_180d}$ & 1.98e+10 & 1.98e+10 & 1e-07 & 3519 & 2381 & 0.25 & 0.52 \\
$\text{UK}^6\text{\_180d}$ & 3.11e+10 & 3.11e+10 & 2e-06 & 5347 & 3888 & 0.21 & 0.41 \\
$\text{Bangalore}^1\text{\_181d}$ & 2.78e+08 & 2.78e+08 & 2e-08 & 3818 & 1360 & 0.33 & 0.84 \\
$\text{Bangalore}^2\text{\_181d}$ & 1.64e+08 & - & - & 5108 & - & 0.24 & - \\
\hline
\end{tabular}
    \caption{Comparison of objective value, solution time and fraction of non-zeros of basic feasible (basic) and interior (inter) solutions of various Calliope models using barrier+crossover (basic) and barrier only (inter), respectively. $\varepsilon$ is the relative error of the interior solution compared with the basic feasible solution. Solution time is in seconds. Non-zeros denotes the fraction of decision variables with absolute value $> 10^{-10}$.}
    \label{tab:inter_vs_basic}
\end{table}

As expected, we can observe that the fraction of non-zeros is significantly higher in interior solutions: In many cases the interior solution contains twice as many non-zero decision variables as the basic feasible solution and almost all interior solutions have more than half of their decision variables away from their bounds. 
Table \ref{tab:nonzeros} breaks down the non-zeros in the $\text{Bangalore}^1\text{\_181d}$ model by decision variable. 
Decision variables in Calliope can have multiple dimensions, e.g. the decision variable energy\_cap is a vector of installed technology capacities with one entry for each technology and location combination. 
Table \ref{tab:nonzeros} highlights why a basic feasible solution might be "nicer" than an interior solution: the interior solution allocates many more technologies that produce and store electricity  (energy\_cap, storage\_cap $> 0$). 
Also it schedules "facilities" to consume and produce energy on more different timesteps (carrier\_prod, carrier\_con $> 0$). 
Clearly, both interior and basic feasible solutions achieve the necessary allocation of carriers at (almost) identical cost, but the actual energy system configurations they imply are quite different.

\begin{table}[ht!]
\centering 
\footnotesize
    \begin{tabular}{|l|r|l|l|p{9cm}|}
    \hline
    \multicolumn{2}{|c|}{}  & \multicolumn{2}{c|}{\textbf{Non-zeros}} & \\
    \hline
\textbf{Variable} & $\textbf{length}$ & \textbf{basic} & \textbf{inter} & \textbf{Explanation} \\
\hline
energy\_cap & 166 & 0.53 & 0.93 & Allocated energy capacity of each technology and location \\ 
carrier\_prod & 628992 & 0.31 & 0.89 & Production of a carrier per location, technology and timestep \\ 
carrier\_con & 567840 & 0.37 & 1.0 & Consumption of a carrier per location, technology and timestep \\ 
cost & 288 & 0.46 & 0.92 & Cost of each cost class per location and technology \\ 
resource\_area & 11 & 1.0 & 1.0 & The area allocated for each technology per location \\ 
storage\_cap & 13 & 0.92 & 1.0 & The storage capacity allocated of each storage technology per location \\ 
storage & 56784 & 0.62 & 1.0 & The energy stored per storage technology, location and timestep \\ 
resource\_con & 48048 & 0.51 & 0.51 & The resource consumption per technology, location and timestep \\ 
resource\_cap & 11 & 1.0 & 1.0 & The resource consumption capacity allocated per location and technology \\ 
carrier\_export & 56784 & 0.0 & 0.58 & The exported amount of a carrier per location, technology and timestep \\ 
cost\_var & 314496 & 0.33 & 0.47 & The variable cost per cost class, technology, location and timestep \\ 
cost\_invest & 288 & 0.42 & 0.88 & The investment cost per cost class, location and technology \\ 
\hline
    \end{tabular}
    \caption{Fraction of non-zero elements in the decision variables of the interior and basic feasible solutions of the $\text{Bangalore}^1\text{\_181d}$. length is the number of entries of the decision variable, the fraction of non-zero elements for the basic feasible solution and the interior solution are indicated as \textit{basic} and \textit{inter}, respectively}
    \label{tab:nonzeros}
\end{table}

\begin{figure}
\centering
  \includegraphics[width=\textwidth]{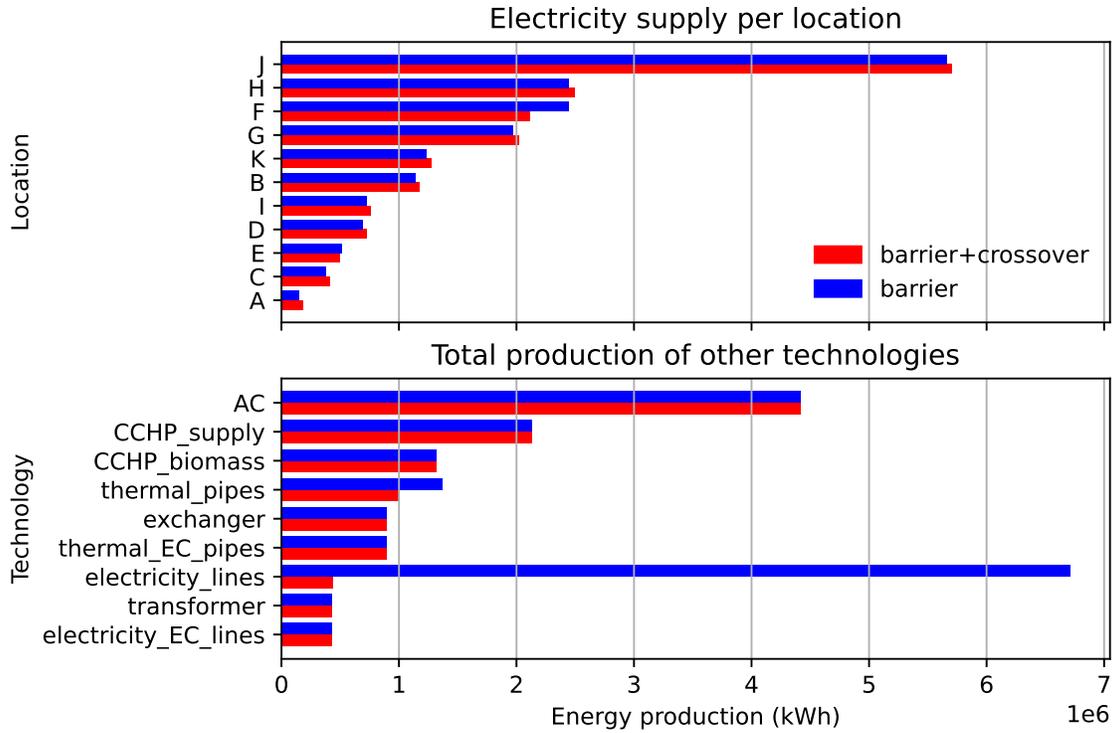}
  \caption{Carrier production \textit{prod} in the $\text{Bangalore}^1\text{\_181d}$ model as computed by barrier+crossover and barrier alone, respectively.
  Top: Electricity supply per location, aggregated over time.
  Bottom: Carrier production of all other technologies, aggregated over time and locations.}
  \label{fig:bangalorehist}
\end{figure}

As discussed in section \ref{subsec_methods}, the most central variables of Calliope energy system models are the installed technology capacities (\textit{cap}) and the energy "production" amounts (\textit{prod}), in addition to the cost parameters which directly and indirectly control these two variables. 
Figure \ref{fig:bangalorehist} shows \textit{prod} of the interior and basic feasible solution of the $\text{Bangalore}^1\text{\_181d}$ model side-by-side. More precisely, it shows the production at "facilities" summed up over all time steps. Figure \ref{fig:bangalorehist} suggests an intuitive interpretation for the differences between the interior and basic feasible solution: In the interior solution, more electricity is produced centrally at location F and then distributed to many other locations using electricity lines. These other locations, in turn, produce less energy themselves. This result arises because the cost of capacity of electricity lines is very low and their usage is free. While a basic solution has as many variables as possible set to zero and thus allocates and uses fewer technologies, an interior solution will in general have all variables away from zero, which do not have an associated cost. In this example, electricity lines are excessively used in the interior solution, because they incur no cost.

In this case the basic feasible solution is arguably more interpretable and more realistic than the interior solution. 
Likely, this effect is pronounced because of the degree of degeneracy of the model we chose: solutions with almost-optimal cost can be very different.
% The optimal technology allocation and carrier production in the interior solution is obfuscated by a lot of "unnecessary" electricity line network use.   

\subsection{Automatically improving scaling}
The major indicators of numerical problems we find across all three examined models are:
\begin{enumerate}
    \item Crossover makes very slow progress and sometimes needs to be restarted.
    \item Simplex makes only very slow progress on the objective value or jumps wildly in the solution space.
    \item Barrier returns a sub-optimal objective.
    \item The solver detects numerical issues and tries to counteract them by: switching to higher precision, dropping variables from the current basis, tightening Markowitz tolerance.
\end{enumerate}
(1) and (2) often occur for the Euro model. (3) is common for the Bangalore model. We conclude that these models exhibit numerical issues that need to be addressed. The UK model, on the other hand, seldom shows any of these issues.

\begin{figure}[ht!]
    \centering
    \includegraphics[scale=0.7]{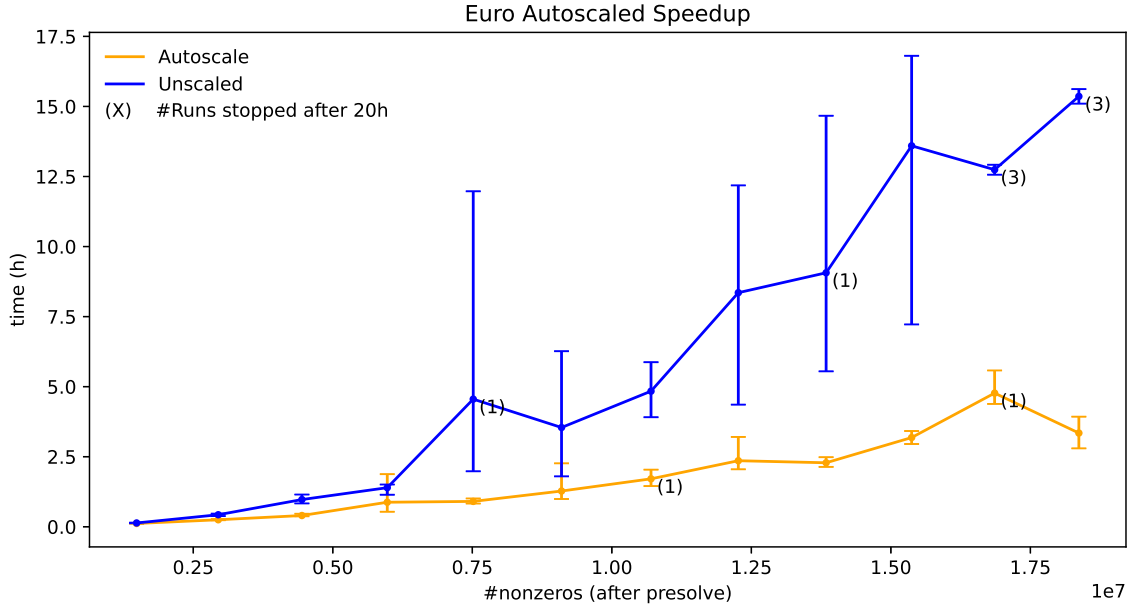}
    \caption{Average absolute solution time for the Euro model for time frames between 1 and 12 months both with and without autoscaling and using barrier+crossover. Average was taken over 5 runs, the vertical bars show the max and min solution time. Runs taking more than 20h were aborted and not considered in either average or errorbars. The number of aborted runs per instance is indicated in brackets next to the average solution time.}
    \label{fig_euro_autoscaling}
\end{figure}

We find that our autoscaling approach can significantly reduce the solution time of the barrier+crossover method for numerically difficult problems. Figure \ref{fig_euro_autoscaling} shows the average solution time of the Euro model for different time frames. Scaling considerably reduces the solution time. Runs that did not successfully converge were not included in the average solution time but instead, their number is indicated in brackets. We notice that autoscaling not only reduces average solution time, but also leads to more regular solution time behaviour and reduces the number of times the algorithm does not converge. However, there are two large outliers in the scaled solution times that did not converge in time. This behaviour seems surprising because in both cases the remaining four runs terminated quickly and with similar solution time. We recall that only the random seed changes within a set of $5$ runs of the same model instance. We will discuss this issue of extreme outliers due to scaling below.

\begin{figure}[ht!]
    \centering
    \includegraphics[scale=0.8]{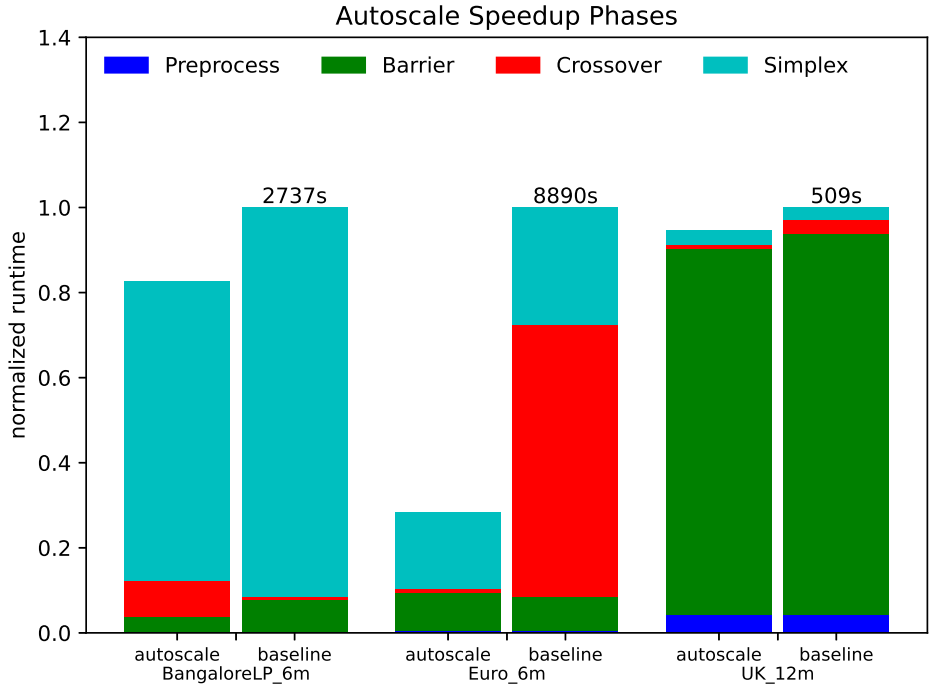}
    \caption{Normalized solution time of barrier+crossover for three models with autoscaling (left bar) and without autoscaling (right bar). The solution time is broken down in the four main phases of the algorithm. Each bar is the average of the solution times of 10 experiments. Absolute solution time in seconds is indicated above the bars of the unscaled models. Error bars were omited to improve readability. }
    \label{fig_autoscaling}
\end{figure}

Figure \ref{fig_autoscaling} shows the effect of autoscaling on Euro\_6m, UK\_12m and BangaloreLP\_6m. We solve each model using Gurobi's barrier+crossover method both with, and without autoscaling enabled. In the Euro\_6m model the solution time improves by about 3x when applying autoscaling. In BangaloreLP there is still a noticeable improvement of solution time after scaling. In the UK model the solution time is hardly affected. This supports our hypothesis that the UK model is generally well formulated already and does not cause numerical issues.
Looking at the numerical range before and after scaling of the three model instances in Table \ref{tab:scaling_range} shows that the original Euro\_6m model has a numerical range close to $16$ orders of magnitude. As discussed in Section \ref{sec_scaling} this may very well be the cause of grave numerical issues. Moreover, scaling achieves a significant reduction in the range of the Euro\_6m model. The other two model instances, however, have a more moderate numerical range and scaling achieves a much smaller improvement of the numerical range which explains the solution time for solving these problems is less affected. Gurobi recommends numerical ranges of at most $10^9$, further supporting these conclusions.

\begin{table}[ht!]
\centering 
\begin{tabular}{|l|r|r|r|r|}
\hline
Model & $\kappa'$ before scaling & $\kappa'$ after scaling  \\
\hline 
Euro\_6m & $3.7\cdot 10^{15}$ & $3.3 \cdot 10^{8}$ \\
UK\_12m & $3.4 \cdot 10^{11}$ & $7.7 \cdot 10^{7}$ \\
BangaloreLP\_6m & $5.1 \cdot 10^{10}$ & $3.1 \cdot 10^{9}$ \\
\hline
\end{tabular}
\caption{Improvement of $\kappa'$ by scaling}
\label{tab:scaling_range}
\end{table} 

In Figure \ref{fig_autoscaling} we report the fraction of the solution time spent in each phase of the barrier algorithm (to be precise, the different sections of each bar represent the fraction between the average of the corresponding phase and the average of the total solution time). 
We split the crossover phase into its two parts: crossover,where a basic point (a vertex) close to the interior solution is found, and simplex, where this basic point is re-optimized using simplex steps.
Interestingly, for the Euro and Bangalore models different phases of the algorithm are affected by scaling: Euro\_6m gains most in the crossover phase (red) whereas for BangaloreLP\_6m the crossover phase with scaling takes even longer than without scaling. BangaloreLP\_6m instead gains most in the simplex phase (cyan). 
The results also suggest that the barrier method is generally least sensitive to scaling: whereas crossover and simplex phases often speed up noticeably, autoscaling usually shows only very minor improvement.

An important insight is that every phase of the algorithm can encounter numerical difficulties and that the exact nature of the problems leading to deteriorating solution time is subtle. In particular, a measure that improves the solution time of some phase may badly affect the solution time of another phase. Finding a scaling that always works and never deteriorates solution time proved to be a major difficulty in designing a suitable autoscaling approach.

\subsection{Performance variability caused by scaling}

As seen above, while scaling often reduces average time to solve a specific model using the barrier+crossover method, it seems to increase the probability of experiencing extremely long solution times for solving certain models (consider for instance the outlier behaviour in Figure \ref{fig_euro_autoscaling}). In particular, we regularly observe that the final simplex phase does not converge. As described above, there are two different flavours of this: either simplex just stops making progress or it jumps wildly in the solution space.
The latter seems to be a strategy Gurobi applies to deal with stalling progress in the simplex solver.

An intuitive explanation for why scaling might induce this behaviour is given by \cite{ref_scaling_12} and briefly discussed in Section \ref{sec_scaling}: while scaling can improve the average condition number, it can increase the condition number of certain bases in the matrix. 
\cite{ref_scaling_12} gives an illustrative example where scaling increases the condition number of all relevant bases in an LP by an arbitrarily large amount. 
As the bases encountered during simplex are subject to randomness, this may explain why scaling in some cases leads to very long solution times or even non-convergence.

In the following, we want to quantify this effect.
Pragmatically, to judge if autoscaling is still useful we want to answer the following questions
\begin{enumerate}
\item What is the average time it takes to solve a given model?
\item How often will we encounter outlier cases with very long solution time or timeout?
\end{enumerate}
We have seen above that scaling can effectively improve average solution time of a model in a previous paragraph.
In order to address the second question we perform the following experiment:
We consider the Euro\_6m model, which exhibits the outlier behaviour discussed above, and run it 100 times with scaling and 100 times without scaling.
We then evaluate the empirical distribution function of these solution times.

\begin{figure}
  \includegraphics[width=\textwidth]{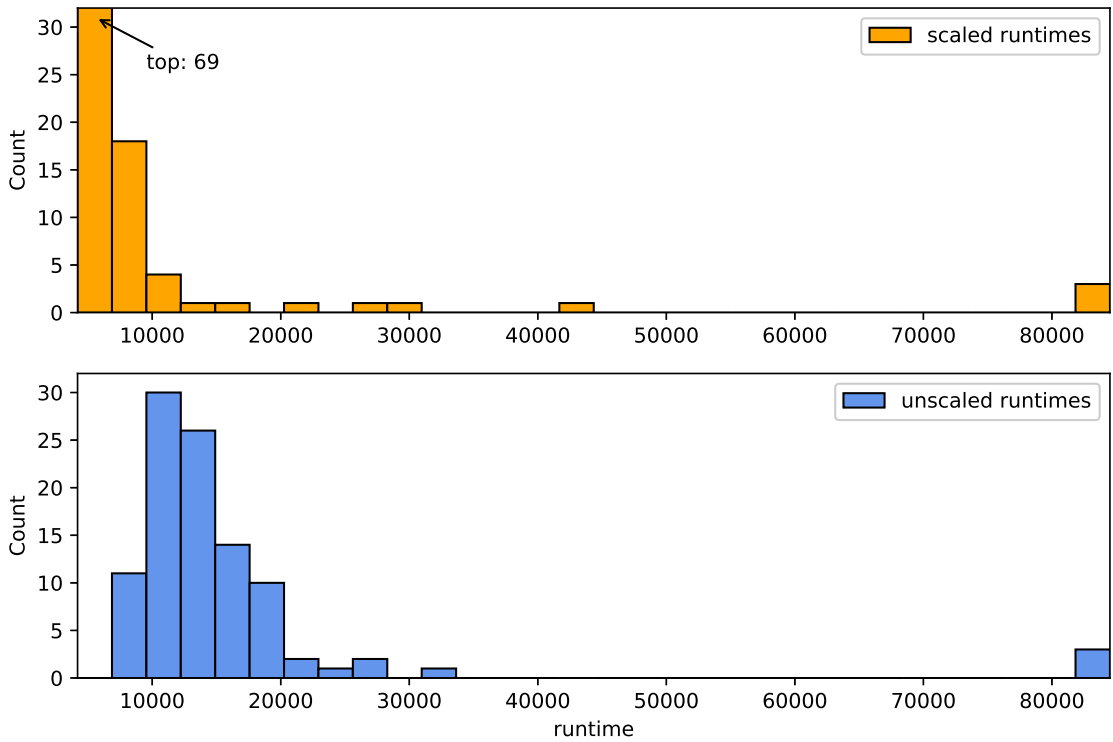}
  \caption{Histogram with 30 bins showing runtimes of 100 runs of each, the scaled and unscaled Euro\_6m model .}
  \label{fig:scaling_distr}
\end{figure}

Figure \ref{fig:scaling_distr} shows the histogram of runtimes of both the scaled and the unscaled data. The runtimes at the right end correspond to timeout runs. While the runtimes of the scaled model have a 40\% lower mean than the ones from the unscaled model, their variance is 25\% higher.
%While the scaled data has a lower mean runtime of 9532s than the unscaled data, which has mean runtime of 15896s, it has a higher variance $2.0 \cdot 10^8s^2 > 1.6 \cdot 10^8 s^2$.
This supports our hypothesis that scaling increases the probability of ``exceptionally long'' solution times; in other words, a distribution fitted to the scaled solution times would have a heavier tail than a distribution fitted to the unscaled solution times.

%Original content from draft
%Using `BarHomogenous` in Gurobi is good because it reduces numerical instability?

%We show how using interior point methods leads to dramatic performance improvements without any substantial impact on results. We analyse whether that indeed is the case.

%For the typical application of energy system models, the solutions are just fine. Also, with Calliope we have the option to automagically remove very small values from the solution, which may make sense based on your intended application. But use responsibly.

%To crossover or not? Maybe IP + crossover is already better than simplex – but turning off crossover is much faster. Also there are different types of crossover.

%If you run barrier methods but don’t turn off crossover, you might not see much gains from using multiple threads, because crossover is once again a pivoting method that is limited to a single thread. Possible result and recommendation: Turn off crossover and ramp up threads to see FASTNESS AND HAPPINESS.

%We show scaling matters for interior point methods performance. We implement auto-scaling in Calliope. Please do enjoy!

\section{Discussion: guidelines for modelling} \label{sec:discussion}
  
We find that barrier is generally superior to simplex for the kinds of models investigated here, and barrier alone is faster than barrier+crossover as the latter performs strictly additional work. The difference in solution time, however, is unpredictable and can be large or barely noticeable, depending on the model instance. Barrier also sometimes fails to converge. In particular, barrier encounters numerical difficulties in models with unbounded (large), optimal surfaces. Such cases can potentially be remedied if crossover is run after barrier. However, we find it difficult to answer the question of whether or not running crossover is worth the additional effort in finding a basic feasible solution. Interior solutions inherently have a larger number of non-zero values in their decision variables as compared to a basic feasible solution. This may negatively affect the interpretability of results, and it can imply very different real-world system designs than those implied by a basic feasible solution -- especially in models in which very different system designs are close to optimal. For instance, the interior solution of one Bangalore model produced more electricity in a centralized location and distributed it via transmission links compared to the basic feasible solution. Investing in the additional effort of crossover must likely be decided on a case-by-case and model-by-model basis, and underscores once again the general problem with relying on a single, ``optimal" result \cite{DeCarolis_Modelling_2016,Lombardi_Policy_2020,Pedersen_Modeling_2021}.

To enable the use of crossover without excessive solution times, we find that model scaling can be helpful; our autoscaling approach often reduces the average solution time of barrier+crossover significantly. This reduction generally happens in the crossover and simplex phases.
From this, we extrapolate that autoscaling has a negligible effect on the solution time of barrier alone but that it improves the average solution time of the simplex method.
We also find that autoscaling increases the probability of solution time outliers: We see more exceptionally long solution times when using autoscaling. This problem requires further investigation in future work. However, as the autoscaling yields almost always shorter solution times and higher convergence rates than our unscaled base cases, autoscaling seems to be a no-regret solution in terms of solution time for barrier+crossover.
% Surprisingly, we find that autoscaling can increase the probability of non-convergence and of extremely long solution times. In other words, the distribution of solution times of scaled model instances has a fatter tail than that of unscaled instances. This problem requires further investigation in future work.

A separate issue which is not easily addressed with scaling is the problem of cost-free technologies. The model $\text{Bangalore}^2$\_181d consistently failed to converge on a solution using the barrier method (Table \ref{tab:inter_vs_basic}). 
For this model, we found that two small modifications to the original model formulation each lead to successful optimization by barrier: increasing costs for the installed capacity of electricity transmission technologies by 2 orders of magnitude, and upper-bounding the energy capacity of electricity transmission.
Both measures counteract the numerical problems caused by unbounded optimal faces, each in a different way, supporting our hypothesis that these problems are due to unbounded optimal faces. 
It is thus advisable to avoid model formulations with free and unbounded technologies. 
There are, however, cases in the real world where some technologies are virtually free to operate or are free to install given the system scope. 
Electricity transmission is one example: their operation cost can be considered negligible compared to the cost of installation, and sizing them might be outside the scope of the problem altogether. 
In these cases it seems unavoidable to resort to dummy costs, but the modeller should be aware that it is non-trivial to set these dummy costs in a way that represents the real world whilst also being high enough to avoid unbounded optimal faces.

If updates to the algorithm, model scaling, and cost-free technologies do not enable model tractability, adjusting solver tolerances may still help. 
Since solver authors made certain assumptions about the input models they will need to solve when setting tolerances, they may not be optimally set for the numerical range or required accuracy of energy system models.
Choosing the right value for tolerances is hard and usually the underlying problem is not solved by modifying tolerances, thus it is only advisable to do so on rare occasions.
Three key tolerances are \textit{Feasibility}, \textit{Optimality} and \textit{Barrier-Convergence}\footnote{When using crossover, there is also the \textit{Markowitz} tolerance, described in more detail in Appendix \ref{sec:markowitz_tol}, but we found no evidence that it improved performance in our benchmarks.}. They all control how tightly some inequality must be fulfilled.
Tightening \textit{Optimality} and \textit{Barrier-Convergence} tolerances will lead to a better objective value in the solution returned by simplex/barrier+crossover methods and barrier methods, respectively. However, tightening these tolerances too much may lead to longer solution times and (in the case of barrier) to non-convergence.
The \textit{Feasibility} tolerance controls how strictly constraints need to be satisfied for a solution to be feasible. Loosening these tolerances may lead to faster convergence of simplex and barrier+crossover methods. However, this usually does not magically resolve all difficulties with numerically challenging problems. It also requires careful experimentation to ensure results do not violate physical properties of the system being described (e.g., negative stored energy).

In the context of these results, we can formulate the following guidelines to improve the computational performance of typical energy system optimisation models:

\begin{itemize}
    \item On large and difficult models, manually select the barrier method or barrier+crossover method. Prefer the latter if you suspect your model formulation to be numerically unstable or if you want a minimal solution in terms of technology allocations.
    \item Use appropriate units that minimize the model's numerical range or apply an automatic scaling procedure like the one we introduce here to derive them automatically.
    \item Be wary of model formulations with cost-free technologies and dummy costs, as those can dramatically worsen the numerical properties of the model and thus increase solution time
    \item Know the basic solver tolerance settings for your chosen solver and adjust them if necessary; however, this should usually be the very last resort.
\end{itemize}

Ultimately, more systematic work to understand the properties of energy system models could help them provide better decision support, for example by making it possible to rapidly explore large numbers of scenarios or alternative solutions even in models that depict the system with high spatial and temporal detail. Promising avenues are custom solvers that exploit these model properties \cite{rehfeldt2021massively}, or even solvers that exploit the fact that a single optimal solution is not necessarily useful; a range of near-optimal solutions, for example extracted from an interior-point algorithm, could be just as relevant for real-world applications \cite{DeCarolis_Modelling_2016,Lombardi_Policy_2020,Pedersen_Modeling_2021}. To complement this, more systematic work is needed on the difference between basic feasible and interior solutions and the implications of these differences for the resulting energy system designs. While black-box solvers like the ones examined here are continuing to be developed and becoming more powerful, the energy modelling community could reap many practical benefits if such work can improve solution times by multiples or even orders of magnitude, as some of our explorations here suggests may be possible.

\bibliographystyle{IEEEbib}
\bibliography{bibl_conf}

\begin{appendices}

\section{Proof of equivalence between scaled and unscaled LP} \label{sec_appendix_proof}
Let us denote by $R$ and $S$ two diagonal matrices 
\begin{equation}
R = \mathbf{diag}(r_1, \dots, r_n), \qquad S = \mathbf{diag}(s_1, \dots, s_n)
\end{equation}
with $r_i, s_i > 0$. Furthermore, let $A \in \mathbb{R}^{m \times n}$, $b \in \mathbb{R}^m$ and $c \in \mathbb{R}^n$. Let's consider the three polyhedra $P := \{x\ |\ A x \leq b\}$, $P'' := \{x\ |\ RAx \leq Rb\}$ and $P' := \{x\ | RASx \leq Rb\}$.  then the following two LPs are equivalent

\begin{equation}
    \min \{c^\intercal x \ |\ x \in P\}, \qquad
    \min \{c^\intercal S x\ |\ x \in P'\}
\end{equation}
To see this note that $P = P''$ because $R$ is invertible and that for each $x \in \mathbb{R}^n$ holds $x \in P' \Leftrightarrow Sx \in P''$. \qedsymbol

\section{Condition number and numerical range} \label{sec_appendix_condition_number}
As discussed in Section \ref{sec_scaling}, both condition number $\kappa(A)$ and the numerical range of $A$ influence how well simplex and barrier can solve the LP with constraint matrix $A$.
We can affect both notions by row and column scaling. Consider the matrix
\begin{equation} \label{eq_matrix_1}
\begin{bmatrix}
1 & 0 \\ 0 & \varepsilon 
\end{bmatrix}
\end{equation}
which has arbitrarily large condition number and numerical range $\frac{1}{\varepsilon}$. 
Scaling the second row of \ref{eq_matrix_1} with $\frac{1}{\varepsilon}$ yields the identity matrix which has both condition number and numerical range of $1$.

On first sight, it may seem as though minimizing numerical range will always also reduce the condition number.
The next example illustrates that this relationship does not hold in general.
consider the matrix
\begin{equation} 
\begin{bmatrix}
1 & \varepsilon \\ \varepsilon & 1
\end{bmatrix}
\end{equation}
Note that if $\varepsilon = 1$ the numerical range of the matrix is $1$, as small as it can get. At the same time, the  matrix is singular, i.e. its condition number is $\infty$. Making $\varepsilon$ go to $0$ will continually increase the numerical range while decreasing the condition number of the matrix. It follows, that we cannot in general optimize both the condition number and the numerical range  simultaneously.

A second problem with improving the condition number of LPs is that it's not possible to improve the condition number of all bases simultaneously: improving the condition number of one basis may adversely affect the condition number of another basis. 
Since it is unknown a priori which bases of $A$ will be visited in the course of the simplex algorithm it is possible that a scaling method has a negative effect on the bases actually encountered during simplex. \cite{ref_scaling_12} investigates the effects of many mainstream scaling methods on the condition number in a large set of practical linear programs.

\section{Scaling Method} \label{sec_method_scaling}
In the following we describe the automatic scaling method we develop for energy system models formulated with Calliope, which we refer to as \emph{autoscaling}. As discussed in Appendix \ref{sec_appendix_condition_number}, minimizing the condition number of the constraint matrix $A$, apart from being hard, may not be desirable. \cite{ref_scaling_75} distinguishes between:
\begin{itemize}
    \item Optimal scaling methods that minimize a certain metric on the constraint matrix $A$.
    \item Empirical scaling methods, such as equilibration, that are sometimes found to work well in practice but don't provide any guarantees about their result.
\end{itemize}
For optimal scaling methods \cite{ref_scaling_75} lists two possible objectives
\begin{align}
    d &:= \frac{\max_{i,j} A_{i,j}}{\min_{i,j} A_{i,j}} \label{eq_d}\\
    \sigma &:= \sum_{i,j} \log(|A_{i,j}|)^2 \label{eq_sigma}
\end{align}
There are algorithms that approximately solve both \ref{eq_d} \cite{algo_d} and \ref{eq_sigma} \cite{algo_sigma}. We devise our own approximation for \ref{eq_d} that incorporates additional domain knowledge about our models.

Most variables in Calliope are associated with some physical quantity such as energy or costs. 
We call this the \emph{type} of the quantity.
Values types can be of vastly different orders of magnitude. 
This effect is often a side effect representing certain types of quantities (cost, energy, etc.) with standard \emph{units} (dollars, kWh, etc.), regardless of the range of their values. 
Informally, autoscaling automates the choice of suitable units for all types of quantities. 
Therefore, it can decrease the numerical range between different types of quantities but not within the same type. 
Note that changing the unit of a type of quantity is a special case of the scaling considered in Section \ref{sec_scaling} with $R = I$ and $S = \mathbf{diag}(f_{u_i})$ where $u_i$ for $i \in [n]$ is the unit of the i-th decision variable. 

In the context of Calliope, it is necessary to distinguish between \emph{base units} and \emph{derived nits}. The sets $U$ of \emph{base units} and $V$ of \emph{derived units} are related as follows:
\begin{equation} 
    u \in U \Longrightarrow u \in V, \quad
    u \in U \Longrightarrow \frac{1}{u} \in V, \quad
    u_1, u_2 \in U \Longrightarrow \frac{u_1}{u_2} \in V
\end{equation}
This distinction is necessary because Calliope models often feature combinations of base units such as \emph{cost per energy}. Scaling \emph{energy} by a factor $s_e$ and \emph{cost} by a factor $s_c$ we necessarily need to scale \emph{cost per energy} by the factor $\frac{s_c}{s_e}$ in order to retain consistency. Associating a special base unit $\textbf{1} \in U$ with all quantities that do not correspond to a physical quantity allows us to represent all derived units $v \in V$ of Calliope models as a fraction of base units $v = \frac{u_i}{u_j}$.
Let $A_u$ the set of values in the model that have unit $u \in V$. Our goal is to find scaling factors $f_u$ for each unit $u \in V$ that minimize
\begin{equation} \label{eq_kappa}
\kappa' := \frac{\max_{u \in U} \max_{a \in A_u} f_u \cdot a}{\min_{v \in U} \min_{b \in A_v} f_v \cdot b}
\end{equation}
Next, we discuss some considerations when choosing scaling factors and then we explain how we actually compute them.

\paragraph{Scaling factors that tamper with the precision of input values will corrupt the model.}
In order to see the significance of numerical precision, consider the following example:
Consider the inequality $0.1 x \leq 100$. If we scale this inequality by $\frac{1}{3}$ we get the inequality $\frac{1}{30} x \leq \frac{100}{3}$. In mathematics, the second inequality is equivalent to the first one, but in floating point arithmetic it is not. Let us assume for simplicity that we work in a decimal floating point system with $5$ decimal digits precision, then the second inequality becomes $0.03333 x \leq 33.33333$. Note that $100$ was scaled by $0.3333333$ and that $0.1$ was scaled only by $0.3333$. The inequality is thus no longer equivalent to the original inequality; in fact, the second inequality now reduces to $x \leq 1000.0999$. Note that the relative error $\frac{1000.0999 - 1000}{1000}$ is exactly $\frac{0.3333333}{0.3333} - 1$. The relative error thus increases with the numerical range of the inequality. This loss of precision can be avoided if all scaling factors are chosen to be powers of $2$.

\paragraph{Optimal scaling factors may lead to prohibitively large or small values.}
Gurobi recommends that all values in a model be between $10^{-3}$ and $10^6$ \footnote{ https://www.gurobi.com/documentation/9.0/refman/num\_advanced\_user\_scaling.html}. The absolute size of values matters to the solver because it internally uses absolute tolerances to compare values. In particular, having input values smaller than the solvers tolerances means that the solver can no longer distinguish between legitimate values and values arising from round-off errors \cite{guidelines_klotz}.
Ensuring that absolute values stay above a certain threshold sometimes limits how much $\kappa'$ can be decreased. Consider the following example:

Let $u$ and $v$ be the two base units in our model and let's assume that the derived units in our model all have the form $u$, $v$ or $\frac{u}{v}$. Assume the values in our model have the ranges as shown in Table \ref{tab:model_ranges}.
\begin{table} 
    \centering
    \begin{tabular}{|l|r|r|}
    \hline 
         unit & min & max  \\
         \hline 
         $u$ & 1 & 50 \\
         $v$ & 0.001 & 100 \\
         $\frac{u}{v}$ & 0.001 & 100 \\
         \hline
    \end{tabular}
\caption{Example range of values in a model to demonstrate the inability to reduce $\kappa'$ if absolute values are to stay within a recommended range.}
\label{tab:model_ranges}
\end{table}
Here $\kappa' = 10^5$ cannot be decreased any further. Assume that we want to ensure that all values in our model have absolute value at least $0.01$. To achieve this, we need to choose scaling factors $f_v \geq 10$ and $f_u \geq 10 \cdot f_v$. Thus after scaling it holds that $\kappa' \geq \frac{10 \cdot f_v \cdot 50}{f_v \cdot 0.001} = 5\cdot 10^5$ which is larger than the original $\kappa'$. There is thus a trade-off between minimizing $\kappa'$ and ensuring that the values have a sensible absolute size.

\paragraph{Optimal scaling factors}
We now formulate an auxiliary optimization problem for finding scaling factors $f_u$ taking into account the results of the previous discussion:
We wish to find scaling factors $S = \{ f_u\ |\ u \in U \} \subseteq \{2^x | x \in \mathbb{Z}\}$ that are powers of two and that minimize the the numeric range between different types of variables while avoiding to scale any variable below some threshold $L$.
Note that all variables of some column $A_i$ of constraint matrix $A$ have the unit of the $i$-th decision variable. 
Thus denote by $u_i, v_i \in U$ for each $i \in [n]$ the base units satisfying that all variables in column $i$ of $A$ have unit $\frac{u_i}{v_i}$. We thus wish to solve
\begin{align}  
&\min_{S \subseteq \{2^x | x \in \mathbb{Z}\}} 
\max_{\substack{0 \leq i,j \leq m \\ 0 \leq k,l \leq n}} 
\frac{
    \frac{f_{u_k}}{f_{v_k}} 
    A_{i,k}
    }{
    \frac{f_{u_l}}{f_{v_l}} 
    A_{j,l} 
} \\
\text{s.t. } &\frac{f_{u_k}}{f_{v_k}} \cdot A_{i,k} \geq L \qquad \forall k \in \{0,\dots,n\}, \forall i \in \{0, \dots, m\}
\end{align}
We can rephrase this into an integer LP by taking logs of the entries of $A$ and of the scaling factors $f \in S$. 
That is, we define $f_u := 2^{g_u}$ for each $u \in U$, and constrain $g_u \in \mathbb{Z}$ to ensure that all scaling factors are powers of two. 
\begin{align} 
\min \quad &r \nonumber \\
\text{s.t. } &g_{u_k} - g_{v_k} + \log(A_{i,k}) - g_{u_l} + g_{v_l} - \log(A_{j,l}) \leq r \qquad \forall\ i,j,k,l \\
&g_{u_k} - g_{v_k}  \geq \log\left(\frac{L}{A_{i,k}}\right) \qquad \forall k \in \{0,\dots,n\}, \forall i \in \{0, \dots, m\} \\
&g_u \in \mathbb{Z} \qquad \forall u \in U
\end{align}
Note that we don't actually need to consider all entries of $A$. It suffices to include the minimum and the maximum of value of each unit. This simplification considerably reduces the amount of constraints from $n^2m^2$ to $|U|^2$ which is usually very small (below 100). 
Moreover, instead of solving the actual integer LP, we can relax the integrality constraint of $g_u$ and round the resulting variables to the closest integer. 
This will give a 4-approximation of the optimal scaling factors.
In practice we find that the integer LP is solved quite rapidly, thus we keep the integrality constraint in place.

\section{Markowitz tolerance} \label{sec:markowitz_tol}

In computing the LU factorization of a matrix, the \emph{Markowitz tolerance} controls which elements are sufficiently large to be used as pivots. An LU factorization of the constraint matrix $A$ is computed at several different stages of the algorithm: during the simplex algorithm in order to recompute the current basis \cite{suhl_suhl_90, suhl_suhl_93, lu_simplex} and during crossover in order to construct a valid basis from scratch \cite{ref_recovering_basis_bixby}. Choosing a large Markowitz tolerance means that many potential pivots are disregarded in order to ensure a numerically stable basis. 
When solving certain numerically challenging models, the crossover method will often need to restart with an updated value of the Markowitz tolerance.
One might expect that setting a more conservative value for the Markowitz tolerance in the first place will prevent this but we could not consistently verify this in our benchmarks.

\section{Model Variations in Experiments} \label{appendix:model:variations}
In this section we describe the model variations we used in \ref{sec_interior}.

\subsection{Euro Model}
\begin{table}[ht!]
\centering 
\footnotesize
    \begin{tabular}{|l|p{3.5cm}|p{1cm}|p{1cm}|p{1.8cm}|p{1.5cm}|p{1.5cm}|}
    \hline
    \textbf{Model} & \textbf{Battery cost} & \textbf{Zones} & \textbf{CO2 caps} & \textbf{Renewable shares} & \textbf{Hydro reservoir} & \textbf{Data source} \\
    \hline
    Euro\_180d                   & 135\euro/kW,315\euro/kWh & 34 & yes & yes & no & original \\
    $\text{Euro}^1\text{\_180d}$ & 135\euro/kW,315\euro/kWh  & 20 & no & no & yes & averaged \\
    $\text{Euro}^2\text{\_180d}$ & 135\euro/kW,315\euro/kWh  & 20 & yes & yes & no & original \\
    $\text{Euro}^3\text{\_180d}$ & 135\euro/kW,315\euro/kWh  & 1 & yes & yes & no & original \\
    $\text{Euro}^4\text{\_180d}$ & 68\euro/kW,158\euro/kWh & 1 & yes & yes & no & original \\
    $\text{Euro}^5\text{\_180d}$ & 34\euro/kW,79\euro/kWh & 1 & yes & yes & no & original \\
    $\text{Euro}^6\text{\_180d}$ & 135\euro/kW,315\euro/kWh  & 1 & no & yes & no & original \\
    \hline
    \end{tabular}
    \caption{}
    \label{tab:euro:model:variations}
\end{table}

\textbf{Battery cost} controls the cost per energy capacity and the cost per storage capacity of batteries. 

% diw_battery_baseline_cost:
%     techs.battery.costs.monetary.energy_cap: 135140.0 # [1.0 EUR per MW] 
%     techs.battery.costs.monetary.storage_cap: 315320.0 # [1.0 EUR per MWh] 
        
% diw_battery_low_cost:
%     techs.battery.costs.monetary.energy_cap: 67570.0 # [1.0 EUR per MW]
%     techs.battery.costs.monetary.storage_cap: 157660.0 # [1.0 EUR per MWh]
        
% diw_battery_lowest_cost:
%     techs.battery.costs.monetary.energy_cap: 33785.0 # [1.0 EUR per MW]
%     techs.battery.costs.monetary.storage_cap: 78830.0 # [1.0 EUR per MWh]

\textbf{Zones} varies the number of countries in the model. While the baseline model consists of 34 European countries, each making up one zone of the model, the variations consist of 20 countries and 1 country (Germany), respectively.

\textbf{CO2 caps} limits the total amount of co2 produced in each location. The bound ranges from $1.2$ Mt in Cyprus to $184$ Mt in Germany.

\textbf{Renewable shares} requires a certain share of the total electricity consumption of each country to exceed a minimum value. This value varies per country and ranges from 11\% in Luxembourg to 78\% in Portugal.

\textbf{Hydro reservoir} controls whether the hydro reservoir technology can be used to store energy (yes) or not (no).

\textbf{Data source} describes what source was used in the model to control the resource constraints of renewables. The original model contains csv timeseries for each of wind, pv and hydro. In \textbf{average}, these values were simply averaged over the whole timerange.

% tech_groups.wind_onshore.constraints.resource: 0.1
% techs.open_field_pv.constraints.resource: 0.137
% techs.roof_mounted_pv.constraints.resource: 0.01
% techs.wind_offshore.constraints.resource: 0.1
% techs.hydro_run_of_river.constraints.resource: 75
% techs.hydro_reservoir.constraints.resource: 75

\subsection{UK Model}
\begin{table}[ht!]
\centering 
\footnotesize
    \begin{tabular}{|l|l|r|r|l|}
    \hline
    \textbf{Model} & \textbf{Battery cost} & \textbf{Imports} & \textbf{Renewables share} & \textbf{New Nuclear} \\
    \hline
    UK\_180d                    & 140£/kW,109£/kWh & $12.75$ GW & 0\% & no \\
    $\text{UK}^1\text{\_180d}$  & 140£/kW,109£/kWh & $12.75$ GW & 0\% & no \\
    $\text{UK}^2\text{\_180d}$  & 300£/kW,200£/kWh & $0$ GW & 80\% & no \\
    $\text{UK}^3\text{\_180d}$  & 75£/kW,50£/kWh & $25.5$ GW & 40\% & yes \\
    $\text{UK}^4\text{\_180d}$  & 150£/kW,100£/kWh & $12.75$ GW & 50\% & no \\
    $\text{UK}^5\text{\_180d}$  & 75£/kW,50£/kWh & $0$ GW  & 90\% & no \\
    $\text{UK}^6\text{\_180d}$  & 75£/kW,50£/kWh & $0$ GW  & 100\% & no \\
    \hline
    \end{tabular}
    \caption{}
    \label{tab:uk:model:variations}
\end{table}

\textbf{Battery cost} controls the cost per energy capacity and the cost per storage capacity of batteries.

% battery cost
% techs:
%     battery:
%         costs:
%             monetary:
%                 energy_cap: 139.69  # 2016 GBP/kW, low cost scenario {own_assumptions}
%                 storage_cap: 109.13  # 2016 GBP/kWh, low cost scenario {own_assumptions}

%     cost_batt_high:
%         techs.battery.cost.monetary:
%             energy_cap: 300  # 2016 GBP/kW, high cost scenario {own_assumptions}
%             storage_cap: 200  # 2016 GBP/kWh, high cost scenario {own_assumptions}

%     cost_batt_low:
%         techs.battery.cost.monetary:
%             energy_cap: 150  # 2016 GBP/kW, low cost scenario {own_assumptions}
%             storage_cap: 100  # 2016 GBP/kWh, low cost scenario {own_assumptions}

%     cost_batt_breakthrough:
%         techs.battery.cost.monetary:
%             energy_cap: 75  # 2016 GBP/kW, breakthrough cost scenario {own_assumptions}
%             storage_cap: 50  # 2016 GBP/kW, breakthrough cost scenario {own_assumptions}
         
\textbf{Imports} sets the high-voltage energy import per zone. The value shown in the table is aggregated over all zones.

% Z2.techs.hvdc_import.constraints.energy_cap_equals: 1400000  #  {ETYS2016} Norwegian Link 2
% Z6.techs.hvdc_import.constraints.energy_cap_equals: 500000  # {ETYS2016} Moyle
% Z7.techs.hvdc_import.constraints.energy_cap_equals: 1400000  # {ETYS2016} NSN
% Z9.techs.hvdc_import.constraints.energy_cap_equals: 450000  # {ETYS2016} East-West (Wales-Ireland)
% Z11.techs.hvdc_import.constraints.energy_cap_equals: 1400000  # {ETYS2016} Viking
% Z15.techs.hvdc_import.constraints.energy_cap_equals: 5200000  # {ETYS2016} NEMO, IFA, Eleclink, Britned
% Z16.techs.hvdc_import.constraints.energy_cap_equals: 1000000  # {ETYS2016} IFA2
% Z17.techs.hvdc_import.constraints.energy_cap_equals: 1400000  # {ETYS2016} FAB

A \textbf{Renewables share} of X\% forces the combined electricity production of wind\_onshore, wind\_offshore, pv\_rooftop, pv\_utility\_scale and hydro to make up at least X\% of the total electricty.

\textbf{New Nuclear} allows for $17.3$ GW additional energy produced by nuclear technology.

    % new_nuclear:
    %     locations:
    %         Z12.techs.nuclear.constraints.energy_cap_equals: 6226000 (before: 1216000.0)
    %         Z13.techs.nuclear.constraints.energy_cap_equals: 2800000 (before: 0.0)
    %         Z15.techs.nuclear.constraints.energy_cap_equals: 1080000 (before: 1080000.0)
    %         Z17.techs.nuclear.constraints.energy_cap_equals: 4252000 (before: 912000.0)
    %         Z5.techs.nuclear.constraints.energy_cap_equals: 2163000 (before: 2163000.0)
    %         Z7.techs.nuclear.constraints.energy_cap_max: 1208000 (before: 1208000.0)
    %         Z9.techs.nuclear.constraints.energy_cap_max: 8593000 (before: 2406000.0)

\subsection{Bangalore Model}
\begin{table}[ht!]
\centering 
\footnotesize
    \begin{tabular}{|l|p{3cm}|}
    \hline
    \textbf{Model} & \textbf{Cost of carbon}  \\
    \hline
    Bangalore\_181d                   & $1$ \\
    $\text{Bangalore}^1\text{\_181d}$ & $9.49$ \\
    $\text{Bangalore}^2\text{\_181d}$ & $0.365$\\
    \hline
    \end{tabular}
    \caption{}
    \label{tab:bangalore:model:variations}
\end{table}

\textbf{Cost of carbon} is the factor of the cost of emitted carbon in the objective function. 

    % cost_of_carbon_low:
    %     run.objective_options.cost_class: {'monetary': 1, 'carbon': 0.365}
    
    % cost_of_carbon_high:
    %     run.objective_options.cost_class: {'monetary': 1, 'carbon': 9.49}

\end{appendices}

\end{document}